\documentclass{aa_edited}

\usepackage{graphicx,txfonts,booktabs,natbib,mathtools,microtype,amssymb,amsmath,xspace,siunitx}
\usepackage[usenames,dvipsnames]{color} 
\usepackage[normalem]{ulem}
\usepackage[caption=false]{subfig} 
\usepackage{orcidlink}

\usepackage{caption}
\usepackage[inline]{enumitem}
\usepackage[acronym]{glossaries}
\glsdisablehyper
\usepackage{physics}
\usepackage{multirow}
\usepackage{hyperref}
\usepackage{booktabs}
\usepackage[capitalise]{cleveref}
\definecolor{linkcolor}{rgb}{0.0,0.3,0.5}
 \hypersetup{
     colorlinks=true,
     linkcolor=linkcolor,
     filecolor=linkcolor,
     citecolor=linkcolor,      
     urlcolor=linkcolor,
     }

\newcommand{\milan}{Dipartimento di Fisica ``G. Occhialini'', Università degli Studi di Milano-Bicocca, Piazza della Scienza 3, 20126 Milano, Italy}
\newcommand{\infn}{INFN, Sezione di Milano-Bicocca, Piazza della Scienza 3, 20126 Milano, Italy}

\newcommand{\mpa}{Max Planck Institut f\"ur Astrophysik, Karl-Schwarzschild-Str. 1,  85748, Garching, Germany}

\newcommand{\pdet}{p_\mathrm{det}}
\newcommand{\VT}{\mathcal{VT}}
\newcommand{\msun}{\mathrm{M_\odot}}
\newcommand{\Rate}{\mathcal{R}}
\newcommand{\reflb}{\mathrm{ref}}
\newcommand{\zref}{z_\reflb}
\newcommand{\Tobs}{T_\mathrm{obs}}
\newcommand{\lambdaLR}{\lambda_\mathrm{LR}}
\newcommand{\yr}{\mathrm{yr}}

\newcommand{\Fe}{\mathrm{Fe}}

\newacronym{gw}{GW}{gravitational wave}
\newacronym{lvk}{LVK}{LIGO/Virgo/KAGRA}
\newacronym{bh}{BH}{black hole}
\newacronym{sfh}{SFH}{star formation history}
\newacronym{dtd}{DTD}{delay time distribution}
\newacronym{mzb}{$\text{MZbin}$}{mass-redshift bin}
\newacronym{lrt}{LRT}{likelihood-ratio test}
\newacronym{mle}{MLE}{maximum likelihood estimators}
\newacronym{vt}{$\VT$}{spacetime-volume}
\newacronym{S/N}{S/N}{signal-to-noise ratio}

\begin{document}

\title{
Targeting black holes from metal-poor progenitors with next-generation gravitational-wave detectors}
\titlerunning{Targeting black holes from metal-poor progenitors with next-generation gravitational-wave detectors}

\author{
Federico Leto di Priolo\inst{1,3}$\,$\orcidlink{0009-0005-5522-9499},
Martyna Chru{\'s}li{\'n}ska\inst{2,3}$\,$\orcidlink{0000-0002-8901-6994},
Davide Gerosa\inst{1,4}$\,$\orcidlink{0000-0002-0933-3579}
}

\authorrunning{Leto di Priolo et al.}

\institute{\milan \and {European Southern Observatory, Karl-Schwarzschild-Str. 2, 85748 Garching, Germany} \and \mpa \and \infn
\\[3pt] \href{mailto:f.letodipriolo@campus.unimib.it}{f.letodipriolo@campus.unimib.it}    
}

\abstract{
Next-generation gravitational-wave detectors such as the Einstein Telescope and Cosmic Explorer will be able to detect binary black-hole mergers out to the cosmic dawn. Mergers observed in the local Universe represent a mixture of systems formed across the entire cosmic history, spanning a wide range of astrophysical environments. Iron-group elements govern metallicity effects on stellar evolution, making metallicity a key tracer that leaves a strong imprint on the black-hole population.
We introduce the concept of a ``target'' redshift, $z_t$, defined as the epoch at which more than $90\%$ of stars form with metallicity $Z < 0.1\,Z_\odot$. This provides a straightforward way to isolate mergers originating from metal-poor environments. The determination of $z_t$ relies on the reconstruction of the cosmic star formation rate density as a function of iron abundance. This reconstruction is not unique, as it depends on the combination of different empirical scaling relations. Consequently, $z_t$ spans a broad range, from $z_t \sim 4$ to $z_t > 10$, depending on the adopted model variation.
We present a statistical framework that enables rapid tests of astrophysical predictions against forecasted observations from next-generation \acrlong{gw} detectors. By quantifying variations in the binary black-hole merger-rate density between the target redshift and the local Universe, our approach maps evolutionary trends across parameter space and estimates the detection statistics required to distinguish genuine astrophysical variations from statistical fluctuations.
}


\maketitle
\section{Introduction}\label{sec:intro}

Current ground-based \acrfull{gw} detectors LIGO, Virgo, and KAGRA (LVK) constrain the redshift distribution of merging binary \acrfullpl{bh} most strongly up to $z \sim 0.2$; credibility intervals grow rapidly at higher redshifts and become uninformative for $z \gtrsim 1$~\citep{2018ApJ...863L..41F,2020ApJ...896L..32C,2024ApJ...970..128S,2025arXiv250818083T,2025ApJ...994L..52T}. Next-generation facilities such as Einstein Telescope (ET; \citealt{2026JCAP...03..081A}) and Cosmic Explorer (CE; \citealt{2021arXiv210909882E}) will detect binary \acrshort{bh} mergers beyond $z \sim 20$, unveiling a whole new segment of the binary \acrshort{bh} population (though accurately constraining whether binaries truly sit at those redshifts will be somewhat challenging; see \ \citealt{2022ApJ...931L..12N,2023PhRvD.107j1302M}). At those redshifts, the potential existence (and hence detection) of primordial  \acrshortpl{bh} of a cosmological origin presents a tantalizing opportunity for new probes of fundamental physics (e.g.,\ \citealt{2024arXiv240721442I}).

From a stellar-physics perspective, extending the observational reach of the binary \acrshort{bh} population to higher redshifts will be transformative.
The population of \acrshort{gw} sources observed in the local Universe is a mixture of systems formed across the entire cosmic history in different environments. This reflects the wide range of possible time delays between progenitor star formation and the eventual compact-binary merger \citep{2010ApJ...716..615O,2012ApJ...759...52D,2016MNRAS.458.2634M,2016MNRAS.463L..31L,2018MNRAS.480.2011G,2019ApJ...886L...1V,2019MNRAS.482..870E,2019MNRAS.490.3740N,2020MNRAS.499.5941D,2020ApJ...901..137S,2022ApJ...931...17V,2024AnP...53600170C}. Sources detected at a given redshift, $z_t$, originate from progenitor stars formed over a narrower range of earlier redshifts ($> z_t$) and, consequently, from a more constrained set of environments.
Different ``{target redshifts}'' $z_t$ will be required to effectively select populations that formed under specific environmental conditions in \acrshort{gw} data from next-generation detectors.

The key tracer that evolves over cosmic time and leaves strong imprints on the binary \acrshort{bh} merger population is the environment metallicity, which is the focus of this work.
Metallicity in massive stars impacts their mass-loss rates via stellar winds \citep{2003ApJ...591..288H,2011A&A...531A.132V,2014MNRAS.441.3703Z,2016MNRAS.459.3432M,2018MNRAS.474.2959G,2020ApJ...888...76M,2024arXiv241111902C}. Stars in high-metallicity environments can lose a substantial fraction of their initial mass before undergoing core collapse, while stars in low-metallicity environments experience much weaker winds, and therefore retain a significantly larger fraction of their mass throughout their evolution \citep{1987A&A...173..293K,2001A&A...369..574V,2018MNRAS.474.2959G}. This directly translates into the masses of the resulting \acrshortpl{bh}.
The dependence of the maximum \acrshort{bh} mass on metallicity is a robust prediction of stellar evolution models, with more massive remnants being systematically produced in more metal-poor environments \citep{2009MNRAS.395L..71M,2010ApJ...714.1217B,2012ApJ...749...91F,2013MNRAS.429.2298M,2015MNRAS.451.4086S}. Metallicity also influences the opacity and nuclear burning processes inside the stars, which in turn affect its radii at various stages of their evolution, especially during post-main-sequence phases \citep{2020A&A...638A..55K,2022MNRAS.516.5816X,2022MNRAS.512.4116F,2023MNRAS.525..706R}. The complex interplay of metallicity-dependent stellar winds and stellar radii propagates through the evolutionary processes that form merging binary \acrshortpl{bh}.

Despite the large variety of proposed binary \acrshort{bh} formation channels -- ranging from isolated to dynamical scenarios, with various subtypes within each class \citep{2021hgwa.bookE..16M,2022PhR...955....1M} -- the overall trend is a significant increase in binary \acrshort{bh} formation efficiency at lower metallicities \citep{2019MNRAS.482.5012C,2025ApJ...979..209V,2018A&A...619A..77K,2018MNRAS.480.2011G,2019MNRAS.490.3740N}. This inextricably links the \acrshort{gw} event rate, the binary \acrshort{bh} mass distribution, and the redshift distribution.
Disentangling such features from the set of binary \acrshortpl{bh} currently observed by the \acrshort{lvk} interferometers is proving challenging \citep{2024A&A...684A.204R,2025A&A...698A..85L,2025ApJ...994L..52T,2025PhRvD.111f3043H}. We need to observe further out; the question is how much further.

In this paper, we present a statistical argument aimed at quantifying the level of variation in the binary \acrshort{bh} merger rate across mass and redshift that can be inferred with ET and CE, and link it to the astrophysical environments one wishes to probe.
Specifically, we assess how precisely the binary \acrshort{bh} population must be measured at a given target redshift, $z_t$, to determine whether its properties differ from those inferred at $z \sim 0$.
Using state-of-the-art estimates for the metallicity-dependent \acrfull{sfh} calibrated on observations, we illustrate how to make astrophysically informed choices for the target redshift with forecasted binary \acrshort{bh} observations from next-generation \acrshort{gw} detectors.

\section{Goals and methods}

\subsection{Target redshift}\label{sec:zt}
Our goal is to quantify the measurement requirements for a conclusive comparison of binary \acrshort{bh} mergers observed at a target $z_t$ and a reference redshift, $\zref \sim 0$, where $z_t$ is such that the selected subpopulation originates from an environment significantly different (e.g., in metallicity) from the one observed at $\zref$. This comparison aims to isolate the impact of the physical conditions under which the progenitors formed and evolved on the resulting \acrshort{bh} merger population.
Time delays between formation and merger imply that the observable population of compact binaries contains systems formed throughout the entire cosmic history. Selecting sources at a given $z_t$ is therefore the most straightforward way to ensure that the observed mergers originated from stellar progenitors that formed and evolved at redshifts $> z_t$.  In an idealized scenario, different choices of $z_t$ correspond to distinct astrophysical environments, and therefore probe different regimes of stellar physics.

To illustrate this concept with a case study, we focus on the metal-poor regime. For the purpose of this work, we define the target redshift, ${z_t}$, as the redshift at which $> 90\%$ of the stars form with a metallicity of $Z < 0.1\, Z_\odot$, where $Z_\odot$ denotes the solar metallicity. Although our approach does not require making any assumption about how binary \acrshort{bh} properties evolve, the target redshift, $z_t$, depends on how the cosmic star formation rate varies with redshift and metallicity. The latter is subject to considerable uncertainties \citep{2019MNRAS.488.5300C,2022MNRAS.516.5737B,2024AnP...53600170C,2025arXiv251115782C}, which propagate to our $z_t$ selection. In Sec.~\ref{sec:zt_res}, we use the recent estimates by~\cite{2025arXiv251115782C} to infer that $z_t \gtrsim 4$ within current uncertainties, though values as high as $z_t \gtrsim 10$ may be required.

For reference in what follows, it is worth keeping in mind that ET should see $\sim 10^2\, \yr^{-1}$ \acrshort{bh} mergers with  $\sim 10\, \msun$ and $\sim 35\, \msun$  at $z \sim 4$. This estimate is based on~Fig.~3.5 by~\cite{2026JCAP...03..081A}, which uses bins with $\Delta z = 0.2$, $\Delta \log m_1 / \msun = 0.1$, the mass distribution model by~\cite{2023PhRvX..13a1048A}, a Madau-Dickinson \acrshort{sfh} \citep{2014ARA&A..52..415M}, and a $\propto t^{-1}$ delay-time distribution, and considers a single ET instrument in a 2L configuration with an arm length of $15\ \mathrm{km}$. 

\subsection{Detectability and merger rate}

Let us first set the notation for merger rates, which form the basis of our target-redshift analysis. The number of \acrshort{bh} mergers, $\mathcal{N}$, per comoving volume, source-frame time, and component masses is given by
\begin{equation}
\frac{\mathrm{d} \mathcal{N}}
{\mathrm{d}V_c\,
\mathrm{d}t_\mathrm{src}\,
\mathrm{d}m_1\,
\mathrm{d}m_2}
\equiv
\mathcal{R}(z, m_1, m_2)\,.
\end{equation}
Converting to redshift and the detector-frame time gives
\begin{equation}
\frac{\mathrm{d} \mathcal{N}}
{\mathrm{d}z\,
\mathrm{d}t_\mathrm{det}\,
\mathrm{d}m_1\,
\mathrm{d}m_2}
= 
\mathcal{R}(z, m_1, m_2)\;
\frac{\mathrm{d}V_c}{\mathrm{d}z}\;
\frac{1}{1+z}\,,
\end{equation}
and thus
\begin{equation}
\frac{\mathrm{d} \mathcal{N}}
{\mathrm{d}z\,
\mathrm{d}m_1\,
\mathrm{d}m_2}
= 
\mathcal{R}(z,m_1,m_2)\;
\Tobs\;
\frac{\mathrm{d}V_c}{\mathrm{d}z}\;
\frac{1}{1+z}\,,
\end{equation}
where $\Tobs = \int \dd t_\mathrm{det}$ is the total observing time.
The number of detections, $N$, per unit mass and redshift is then given by
\begin{align}
\frac{\mathrm{d} N}{\mathrm{d}z\,\mathrm{d}m_1\,\mathrm{d}m_2}
=
\frac{\mathrm{d} \mathcal{N}}
{\mathrm{d}z\,
\mathrm{d}m_1\,
\mathrm{d}m_2}\;
\pdet(z, m_1, m_2)\,,
\end{align}
where $\pdet$ is the detection probability.
It is convenient to define the differential sensitivity \acrlong{vt},
\begin{align}
\VT(z, m_1, m_2)= 
\frac{\mathrm{d}V_c}{\mathrm{d}z}\;
\frac{\Tobs}{1+z}\;
\pdet(z, m_1, m_2)\,,
\end{align}
which relates to the commonly used definition,
$VT=\int \VT\,\dd z$,
employed to quantify the reach of a GW detector
\citep{2016ApJ...833L...1A,2018CQGra..35n5009T,2020CQGra..37d5007K,2020PhRvD.102j3020G,2021CQGra..38e5010C}.
With this notation, one has
\begin{align}\label{eq:merger_rate_density}
\frac{\mathrm{d} N}{\mathrm{d}z\,\mathrm{d}m_1\,\mathrm{d}m_2}
=
\mathcal{R}(z, m_1, m_2)\,
\VT(z, m_1, m_2)\,.
\end{align}

We estimated \acrshort{gw} detectability by thresholding the \acrfull{S/N} $\rho$. The detection probability is given by
\begin{equation}
\pdet(z, m_1, m_2) = \int H[\rho(z, m_1, m_2,\theta) - \rho_\mathrm{thr}] p(\theta) \dd\theta\,,
\end{equation}
where $H$ is the Heaviside step function, $\theta$ collectively denotes all remaining binary parameters, $p(\theta)$ is their distribution (which we assume is independent of $m_1$, $m_2$, and $z$), and $\rho_\mathrm{thr}$ is the chosen \acrshort{S/N} threshold.
We assumed that binaries are distributed uniformly in sky location, inclination, and polarization. We considered non-precessing sources and modeled their spin magnitudes using the population inferred by the \acrshort{lvk} Collaboration under their fiducial model \citep{2025arXiv250818083T}.
We sampled the spin distribution's hyperparameters from their inferred posterior~\citep{2025arXiv250818083T}
and drew spin values from the resulting distribution. We stress that the detailed black hole (BH) spin treatment is largely irrelevant for our purposes, as spin distributions have only a minor impact on \acrshort{gw} detectability, $\pdet$ -- a factor of a few for \acrshort{lvk}, and up to $\sim 10\%$ for next-generation detectors; see, for example,~\cite{2018PhRvD..98h4036G}.
We computed \acrshortpl{S/N} using the \textsc{IMRPhenomXHM} waveform model \citep{2020PhRvD.102f4002G} and the corresponding routines as provided in the \textsc{gwbench} package \citep{2021CQGra..38q5014B}. We considered a next-generation detector network consisting of a triangular ET interferometer with 10 km arms and a CE interferometer with 40 km. We adopted the sensitivity curves by~\cite{2023JCAP...07..068B} and~\cite{CEcurve}, more specifically those labeled ``ET-10-XYL'' and ``CE-40'' in \textsc{gwbench}.
We set $\rho_\mathrm{thr} = 8 \sqrt{2}$ and used cosmological parameters from~\cite{2020A&A...641A...6P}.

\begin{figure}
    \centering
    \includegraphics[width=\columnwidth]{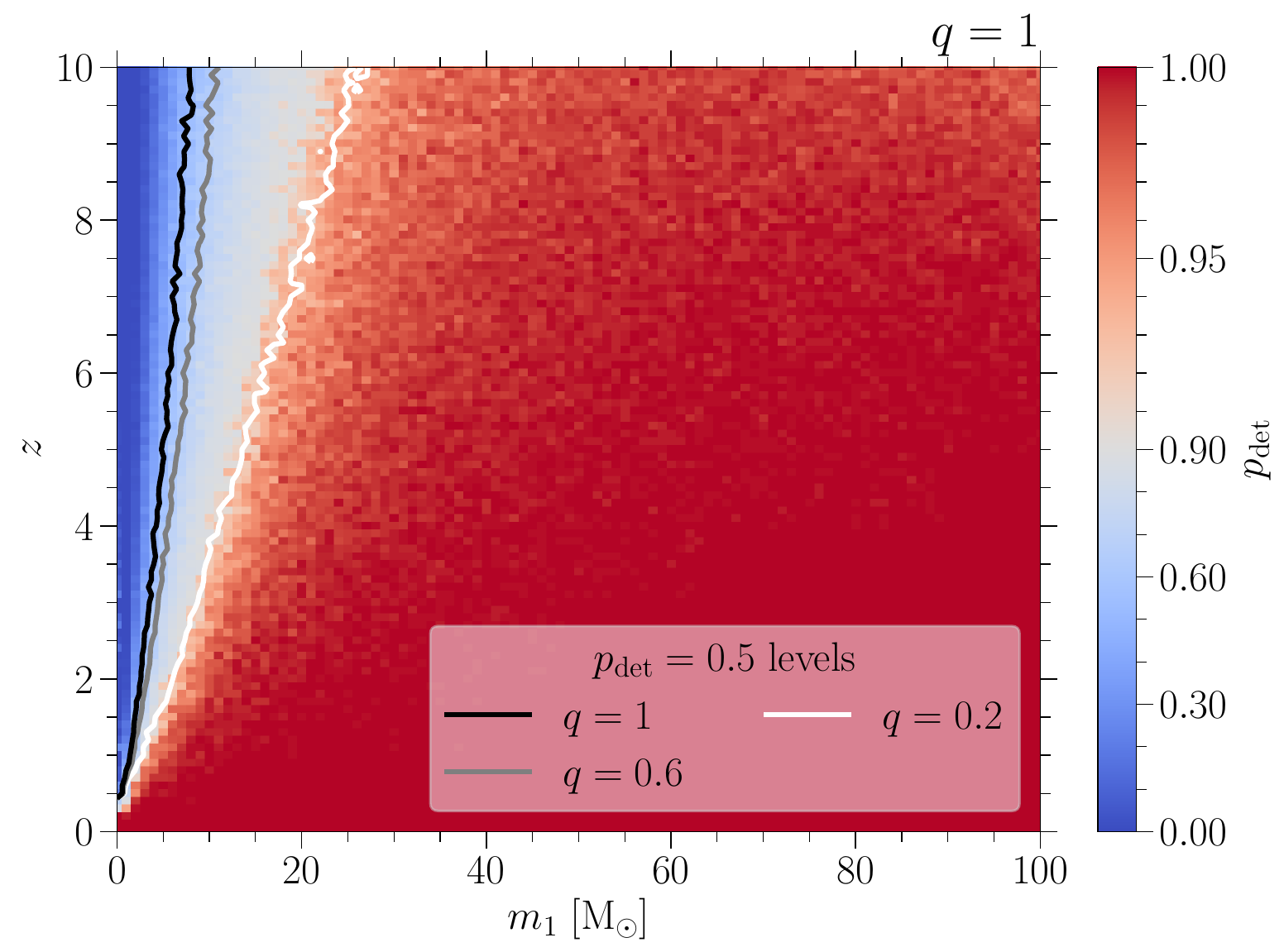}
    \caption{Detection probability for the ET+CE network as a function of primary \acrshort{bh} mass, $m_1$, and redshift, $z$. The color map refers to equal-mass binaries ($q = 1$). Solid contours indicate $\pdet = 0.5$ for different values of the mass ratio.
    }\label{fig:pdet_map}
\end{figure}

\begin{figure}
    \centering
    \includegraphics[width=\columnwidth]{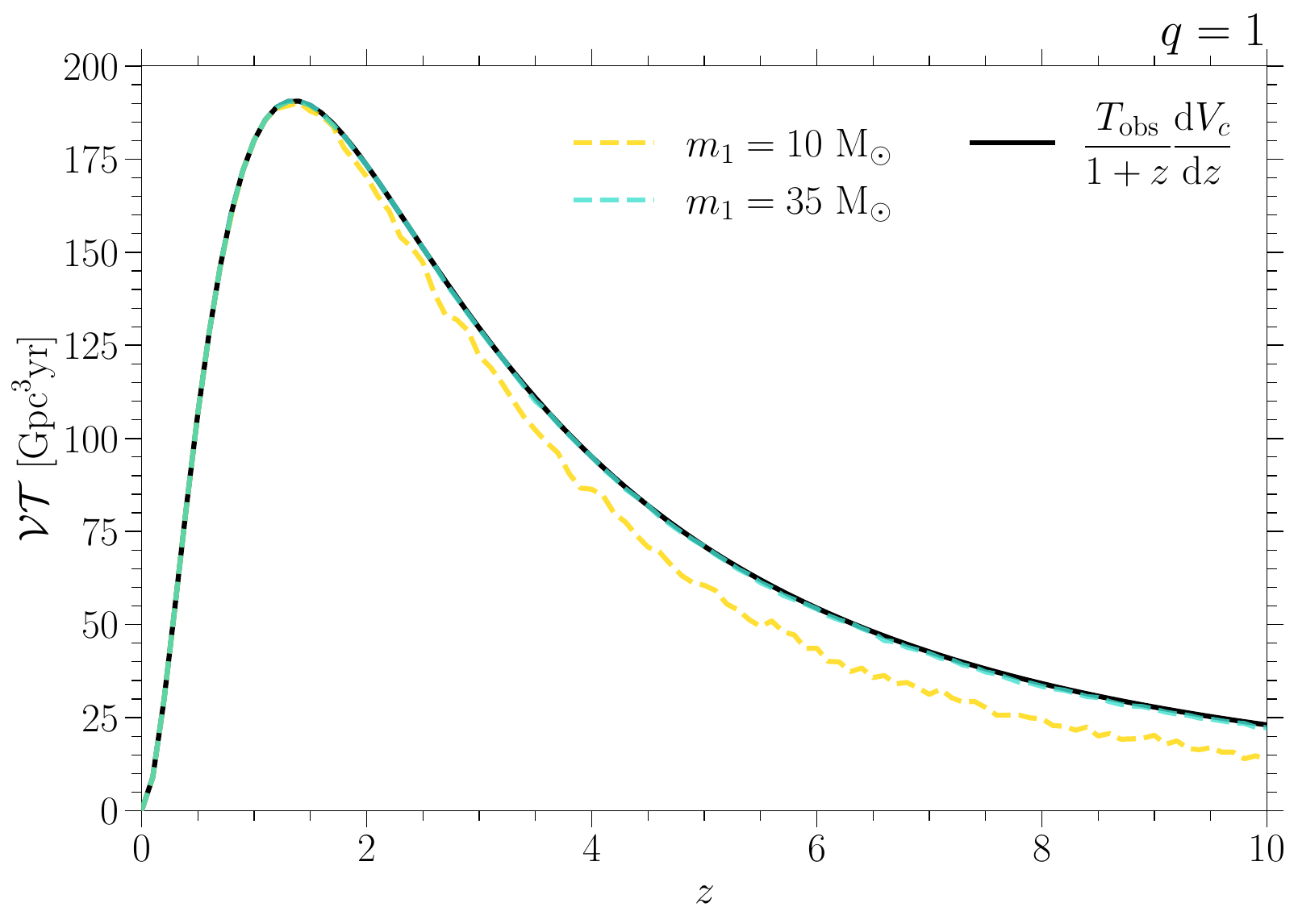}
    \caption{Evolution of the differential spacetime-volume, \acrshort{vt}, with redshift for different binary \acrshort{bh} masses, assuming equal-mass sources ($q = 1$) and $\Tobs = 1\, \yr$. The solid curve assumes $\pdet = 1$ for all sources; the colored curves correspond to \acrshort{bh} binaries with $m_1 = 10\, \msun$ and $35\, \msun$.}\label{fig:pdet_and_VT}
\end{figure}

For a given $(z, m_1, m_2)$ triplet, we simulated a population of $500$ sources over the remaining parameters, $\theta$, and estimated the detection probability, $\pdet(z, m_1, m_2)$, as the fraction of sources yielding a \acrshort{S/N} above the threshold. Detectability was evaluated on a $z$--$m_1$--$m_2$ grid with $\sim 3.5 \times 10^6$ nodes and interpolated using the \textsc{RegularGridInterpolator} linear scheme from \textsc{SciPy} \citep{2020NatMe..17..261V}. We excised regions of parameter space with detection probabilities greater than $0.999$ (smaller than $0.001$) and approximated those values of $\pdet$ as $1$ ($0$). \Cref{fig:pdet_map} shows the resulting detectability across the mass-redshift plane. Interpolation residuals are $\lesssim 5\%$.

\Cref{fig:pdet_and_VT} illustrates how $\VT$ varies with mass and redshift, assuming equal-mass binaries and $\Tobs = 1\, \yr$. The solid line corresponds to $\VT$ computed assuming $\pdet = 1$ for all sources. This limit does not depend on the chosen detector network.
For next-generation detectors and up to $z = 10$, this term largely dominates the estimate of the merger rate. The detectability, $\pdet$, provides a subdominant correction for masses $m_1 \sim 35\, \msun$. For masses of $\sim 10\, \msun$, on the other hand, $\pdet$ drops to $\sim 0.6$ by redshift $z = 10$, providing a more substantial correction to $\VT$.

\subsection{Merger density ratio}

Our framework relies on inferring the relative merger-rate density of binary \acrshortpl{bh} between two bins, as a function of mass and redshift, given a set of \acrshort{gw} measurements. Let us consider two mass-redshift bins, which we refer to as the ``target'' and ``reference'' bins. If the bin widths are much smaller than the extent of the targeted parameter space, the expected number of detections in a given bin can be approximated as
\begin{equation}
N
=
\mathcal{R}(z, m_1, m_2)\,
\VT(z, m_1, m_2)\,
\Delta z\, \Delta m_1\, \Delta m_2\,.
\end{equation}
Assuming bins of equal width, the ratio of the merger rates between the target and reference bins is then given by
\begin{equation}\label{eq:Rate_ratio}
     \frac{\mathcal{R}_t}{\mathcal{R}_\reflb}
     =
     \qty(\frac{N_t}{N_\reflb})
     \times
     \qty(\frac{\VT_\reflb}{\VT_t})\,.
\end{equation}
The first term on the right-hand side in the equation above reflects the observed number of detections, while the second term encodes selection effects and corrects for the different sensitivity of the instrument to sources in the two bins.

Estimating the merger rate density ratio between two bins from the number of detections they contain requires one to account for both the intrinsic Poisson uncertainty associated with a counting process and the different detector sensitivities to each bin, encoded in \acrshort{vt}. We developed two approaches to treat the Poisson uncertainty and subsequently extended them to include the \acrshort{vt} dependence. These two methods are based on likelihood ratios and Bayesian inference, respectively, and are described in the following sections. While results in Sec.~\ref{sec:results} are presented using the Bayesian approach, it is informative to briefly outline the likelihood method as well.
Some properties of relevant probability distributions are summarized in Appendix~\ref{sec:apx_bayes}. Details of the likelihood-based calculations are provided in Appendix~\ref{sec:apx_lrt}, while Appendix~\ref{sec:apx_LRT_vs_Bayes} presents a comparison between the likelihood-based and Bayesian approaches.

\subsubsection{Likelihood-ratio test}\label{sec:lrt}

The \acrlong{lrt} is a hypothesis test that compares the likelihoods of two nested models: a full model and a reduced model that represents a restricted version of the former. The restrictions imposed on the likelihood parameters define the null hypothesis, $\mathcal{H}_0$. 
The test proceeds by computing the \acrlong{mle} of the parameters under $\mathcal{H}_0$ and $\mathcal{H}_1$, denoted by $\hat{\theta}_0$ and $\hat{\theta}_1$, respectively, and evaluating the test statistic \citep{10.5555/42427}
\[
\lambdaLR = -2 \qty[\ln \mathcal{L}(\hat{\theta}_0) - \ln \mathcal{L}(\hat{\theta}_1)]\,.
\]
Under $\mathcal{H}_0$, the statistic $\lambdaLR$ asymptotically follows a $\chi^2$ distribution \citep{Wilks1938TheLD}.

We assumed that the likelihood of measuring $N_t$ events at the target redshift and $N_\reflb$ at the reference redshift is the product of two Poisson distributions with expectation values $\lambda_t$ and $\lambda_\reflb$:
\begin{equation}
\mathcal{L}(\lambda_t, \lambda_\reflb) =
\mathrm{Poisson}(\lambda_t)\,\mathrm{Poisson}(\lambda_\reflb)\,.
\end{equation}
We reparametrized the likelihood in terms of $\lambda_\reflb$ and
$\alpha = \lambda_t / \lambda_\reflb$, and computed the test statistic for the null hypothesis $\alpha = \alpha_0$. This yields
\begin{equation}\label{eq:lambda_lr_log}
    \lambdaLR = 2 N_\reflb \qty[\alpha\ln\qty(\dfrac{\alpha}{\alpha_0}) - (1 + \alpha)\ln\qty(\dfrac{1 + \alpha}{1 + \alpha_0})]\,;
\end{equation}
the derivation is provided in Appendix~\ref{sec:apx_lrt}.
The likelihood has two parameters, one of which is fixed under the null hypothesis. Therefore, $\lambdaLR$ is asymptotically distributed as a $\chi^2$ with one degree of freedom. The 95\% confidence interval for $\alpha = \lambda_t / \lambda_\reflb$ is then obtained by finding all values of $\alpha_0$ that are not rejected by the test. In practice, given measurements of $\alpha$ and $N_\reflb$, the boundaries of the interval are obtained by solving~\cref{eq:lambda_lr_log} for $\alpha_0$.

The relation between $\lambda_t / \lambda_\reflb$ and $\Rate_t / \Rate_\reflb$ follows directly. Considering two \acrlongpl{mzb}, the merger rate density ratio between the target and reference bins is given by~\cref{eq:Rate_ratio}. The confidence interval for $\Rate_t / \Rate_\reflb$ can therefore be obtained by multiplying the values of $\alpha_0$ derived above by the factor $\VT_\reflb / \VT_t$, which is determined by the chosen bins.

\subsubsection{Bayesian inference method}\label{sec:bayes}

In Bayesian probability theory, a conjugate prior for a likelihood function is a prior distribution chosen such that the prior and posterior belong to the same family of probability distributions. Working with conjugate priors often greatly simplifies the calculations. The conjugate prior for the rate parameter, $\lambda$, of the Poisson distribution is the gamma distribution \citep{fink1997compendium}. Let the prior on $\lambda$ be $\mathrm{Gamma}(\alpha, \beta)$, where $\alpha$ and $\beta$ are the shape and rate (or inverse scale) parameters of the distribution, respectively. Throughout this work we use an uninformative Jeffreys prior $\propto \lambda^{-1/2}$, which corresponds to $\alpha = 0.5$ and $\beta = 0$. Then, given a sample of $n$ observations, where $k_i$ is the number of events measured in the $i$-th observation ($i = 1, \dots, n$), the posterior distribution is
\begin{equation}\label{eq:Gamma_posterior}
    \lambda \sim \mathrm{Gamma}\qty(\alpha + \sum_{i=1}^n k_i, \beta + n)\,.
\end{equation}
We limited ourselves to a single set of $N$ events, such that $\lambda \sim \mathrm{Gamma}\qty(\alpha + N, \beta + 1)$.

The ratio of two independent gamma-distributed random variables follows a generalized beta prime distribution (\citealt{alma990006448740302711}; see Appendix~\ref{sec:apx_bayes} for details). In symbols,  if $X_k \sim \mathrm{Gamma}(\alpha_k, \beta_k)$  with $k = 1, 2$, then
\begin{equation}\label{eq:betaprime_posterior}
    \dfrac{X_1}{X_2} \sim \beta'\qty(\alpha_1, \alpha_2, 1, \dfrac{\beta_2}{\beta_1})\,.
\end{equation} 
Using property~\labelcref{eq:betaprime_prop1}, applying~\cref{eq:Gamma_posterior,eq:betaprime_posterior} to our target and reference bins, and assuming equal Jeffreys priors for $\lambda_t$ and $\lambda_\reflb$, we find
\begin{equation}\label{eq:Rate_ratio_posterior}
    \dfrac{\Rate_t}{\Rate_\reflb} \sim \beta'\qty(0.5 + N_t, 0.5 + N_\reflb, 1, \dfrac{\VT_\reflb}{\VT_t})\,.
\end{equation}
\Cref{eq:Rate_ratio_posterior} provides a straightforward method to retrieve the posterior distribution of the merger rate density ratio between two fixed \acrlongpl{mzb}, given a set of observations.

\section{Results}\label{sec:results}

\subsection{Selecting metal-poor progenitors}\label{sec:zt_res}

As described in Sec.~\ref{sec:zt}, we tie our target redshift to the epoch at which at least 90\% of cosmic star formation occurred at metallicities an order of magnitude below solar. Such conditions were characteristic of the early Universe, which was dominated by young galaxies that had not yet undergone substantial chemical enrichment through stellar nucleosynthesis. In contrast, star formation in the present-day Universe spans a much broader range of chemical compositions.

\begin{figure}
    \centering
    \includegraphics[width=\columnwidth]{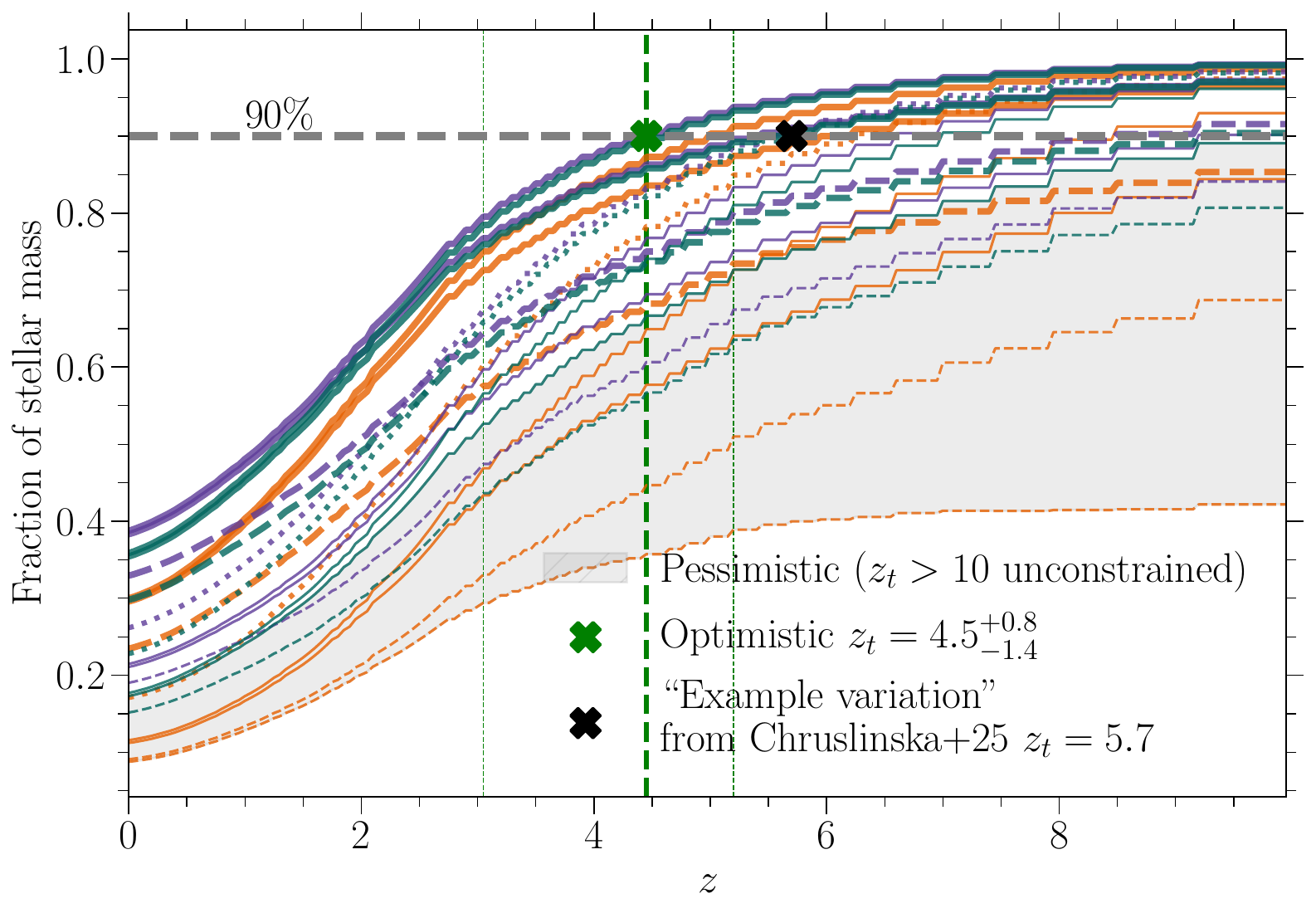}
    \caption{Fraction of stellar mass formed with iron abundance below 10\% of the solar value between redshift, $z$ (horizontal axis), and 10. Different lines correspond to distinct assumptions for the iron-dependent cosmic \acrshort{sfh} (see Sec.~\ref{sec:zt_res}). The target redshift is defined by the intersection of each model curve with the 90\% level. The optimistic scenario yields $z_t \sim 4.5$, whereas in the pessimistic case $z_t\gtrsim 10$.
    Colors indicate the adopted relation between the O/Fe abundance ratio and specific star formation rate of galaxies (purple, orange, and green correspond to the “slow,” “fast,” and “mixed” Fe-enrichment scenarios from \cite{2025arXiv251115782C}, respectively). Solid and dashed lines denote galaxy stellar mass functions with evolving and fixed low-mass slopes, respectively. Thick lines correspond to a shallower slope of the star formation rate-mass relation, while dotted lines allow for additional evolution at $z > 2$ relative to the baseline relation from \cite{2023MNRAS.519.1526P} (thin lines). See text and~\cite{2025arXiv251115782C} for details.}\label{fig:Mf_zt}
\end{figure}

To determine how early in cosmic history these conditions prevailed, and thus to estimate $z_t$, we used the suite of metallicity-dependent cosmic \acrshort{sfh} models by~\cite{2025arXiv251115782C}, who extend and validate the observation-based framework by~\cite{2019MNRAS.488.5300C} and~\cite{2021MNRAS.508.4994C} against more recent empirical constraints. Within this framework, the metallicity-dependent cosmic \acrshort{sfh} is obtained by combining empirical relations linking galaxy properties (star formation rate, gas-phase oxygen abundance, and stellar mass) with their number statistics.
\cite{2025arXiv251115782C} introduce a set of model variations that bracket the dominant uncertainties in these relations and propagate them to the final result. These variations explore different assumptions regarding: (i) the low-mass slope of the galaxy stellar mass function and its redshift evolution; (ii) the slope and redshift evolution of the star formation rate-mass relation; (iii) the normalization and high-redshift ($z > 3$) evolution of the relation connecting galaxy gas-phase oxygen abundances to their properties (masses, star formation rates); and (iv) the form of the relation between the oxygen-to-iron abundance ratio and the specific star formation rate of galaxies.
The latter relation (iv), introduced by \citep{2024A&A...686A.186C}, encapsulates the delay in enhanced iron enrichment by Type~Ia supernovae (SN Ia) relative to the prompt enrichment by core-collapse supernovae from massive progenitors. The uncertain delay time distribution of SN Ia is one of the key factors shaping the relation. \cite{2025arXiv251115782C} apply the relation (iv) to convert gas-phase oxygen abundances into iron abundances and derive the iron-dependent cosmic \acrshort{sfh}\@.
This step is crucial for a meaningful determination of $z_t$, as iron-group elements set the metallicity dependence of stellar evolution by regulating stellar opacities and radiation-driven winds \citep{1996ApJ...464..943I,2001A&A...369..574V,2020MNRAS.499..873S,2025A&A...697A.114P}, and thereby influence the properties of stellar-origin compact objects and related transients \citep{2002RvMP...74.1015W,2006ApJ...638L..63L,2010ApJ...715L.138B,2016ApJ...817....8P,2018A&A...619A..77K,2018MNRAS.473.1258S,2022MNRAS.516.5737B,2023MNRAS.525..706R,2023MNRAS.524..426I,2024AnP...53600170C,2024A&A...688L..10V,2025MNRAS.543.2796H}.
Accordingly, throughout this work, we use the term ``metallicity'' to refer to iron abundance.
\Cref{fig:Mf_zt} illustrates the set of model variations from~\cite{2025arXiv251115782C} in the context of our target-redshift definition. Each curve shows the fraction of stellar mass formed with metallicity below $0.1\,Z_\odot$ between redshifts $z$ and 10, and thus defines a corresponding target redshift via its intersection with the 90\% horizontal line.
Each model results from different combinations of the ingredients (i-iv) used to construct the observation-based, iron-dependent cosmic \acrshort{sfh}.

The model variations in~\cref{fig:Mf_zt} span a wide range of possible target redshifts. We define two representative thresholds, corresponding to the lower and upper bounds on $z_t$ implied by the metallicity-dependent cosmic \acrshort{sfh} models. These are referred to as optimistic and pessimistic, respectively, reflecting the detection statistics required to accurately constrain variations in the merger rate between $z_t$ and the local Universe. 
Even with next-generation facilities, detecting signals from such ``pessimistic'' redshifts does not necessarily imply accurately inferring their redshift; see, e.g.,~\cite{2023PhRvD.107j1302M}. In this sense, lower target redshifts are preferable.
In the optimistic case (indicated by the green cross in~\cref{fig:Mf_zt}), we obtain $z_t$ = 4.5, with a more liberal lower bound of $z_t>3.1$ (vertical green line in~\cref{fig:Mf_zt}) when considering possible systematic uncertainties in (iii) arising from ambiguities in the oxygen abundance scale \citep{2019A&ARv..27....3M,2019ARA&A..57..511K}.
This scenario requires ``slow'' \(\Fe\) enrichment relative to star formation (corresponding to a long minimum SN Ia delay time of 400 Myr for a conventional $\propto t^{-1}$ delay-time distribution) and a significant contribution from low-mass galaxies ($M_{*}\lesssim10^{8}\ M_{\odot}$) to the cosmic star formation rate budget. The latter reflects the importance of such galaxies in shaping the low-metallicity tail of the cosmic \acrshort{sfh}, and in the corresponding model is realized through a steepening of the galaxy stellar mass function toward low masses with increasing redshift, together with a shallower slope of the star formation rate-mass relation (i.e., $\log(\mathrm{SFR}) \propto 0.8 \log(M_{*})$, compared to $\log(\mathrm{SFR}) \propto \log(M_{*})$ scaling inferred from the literature meta-analysis by~\citealt{2023MNRAS.519.1526P} and used as a baseline relation in~\citealt{2025arXiv251115782C}).
In contrast, model variations that minimize the contribution from low-mass galaxies (thin dashed lines) yield higher $z_t$, an effect that is further reinforced when combined with ``fast'' $\Fe$ enrichment scenario (orange lines, corresponding to a minimum SN Ia delay time of 40 Myr). A number of iron-dependent cosmic \acrshort{sfh} estimates derived under such assumptions result in $z_t > 10$, i.e.,\ values that cannot be constrained within the redshift range covered by these models. These cases define the pessimistic scenario and span the shaded gray region in~\cref{fig:Mf_zt}.
We also mark $z_t = 5.7$, obtained for a representative model combining more intermediate assumptions (the ``example variation'' from~\cite{2025arXiv251115782C}, black cross), in particular, assuming a ``mixed'' Fe enrichment from multiple Type~Ia supernova progenitor channels with a delay-time distribution from~\cite{2005A&A...441.1055G}.  
Even in this intermediate case, selecting metal-poor progenitors according to our condition requires one to target the first $\lesssim 1\,\mathrm{Gyr}$ of the Universe’s history.

\subsection{Targeted features in the mass spectrum}\label{sec:mass_features}
We focus our discussion on two bins representing typical ``low-mass'' ($\sim 10\,\msun$) and ``high-mass'' ($\sim 35\,\msun$) \acrshort{bh}s in systems merging at low redshift, motivated by the observed features in the mass spectrum identified by \acrshort{lvk} \citep{2025arXiv250818083T,2026arXiv260527226T,2025arXiv250915646B,2026ApJ...996..144G}.
The peak in the primary mass distribution at around  $10\,\msun$  is currently the most robustly detected feature.
A broader overdensity (with a merger rate density of $\sim 1/50$ of that measured in the low-mass peak) is also present at $\sim 35\,\msun$, though the statistical significance of this feature is model-dependent  \citep{2024ApJ...962...69F,2025PhRvD.111f1305H,2025arXiv251122093S,2025ApJ...994L..52T} and may instead reflect fluctuations due to limited sample sizes \citep{2026arXiv260300239C}. At higher masses the BH merger rate density declines more steeply.

Stellar evolution, binary interactions, and metallicity all imprint on the binary \acrshort{bh} merger mass distribution, complicating the interpretation of these features and shaping their evolution with redshift \citep[e.g.,][]{2025arXiv250818083T}. 
The low-mass peak is predominantly attributed to BH mergers formed through the evolution of isolated binaries, via both common-envelope evolution and stable mass transfer channels \citep[e.g.,][]{2022PhR...955....1M}, but may also include a contribution from dynamically assembled mergers in dense stellar clusters \citep[e.g.,][]{2026ApJ...997..267Y}.

The origin of the high-mass overdensity is heavily debated. The appealing interpretation as a pileup caused by (pulsational) pair-instability supernovae \citep{2015A&A...573A..18M,2016MNRAS.457..351Y,2019ApJ...882...36M,2019ApJ...882..121S,2021ApJ...912L..31W}, which can map stellar progenitors spanning a wide range of masses onto a relatively narrow range of BH masses, is difficult to reconcile with current models, which generally predict such a feature at higher masses \citep[e.g.,][]{2023MNRAS.526.4130H}. Multiple formation channels, including isolated binaries evolving through stable mass transfer \citep[e.g.,][]{2025arXiv251220054X} or chemically homogeneous evolution \citep[e.g.,][]{2016MNRAS.460.3545D}, as well as dynamically assembled systems and hierarchical mergers in dense stellar environments \citep[e.g.,][]{2025MNRAS.538..639B}, can all contribute to the formation of BH mergers in this mass range.

The low- and high-mass BH mergers likely arise from different (combinations of) formation channels, each characterized by distinct metallicity dependencies and delay time distributions. For example, the stable mass-transfer channel typically yields longer delay times than the common-envelope channel \citep{2021ApJ...922..110G,2021A&A...645A..54K,2026A&A...706A.296K,2021A&A...650A.107M,2022ApJ...931...17V}, and both channels may become inefficient at high metallicity. Chemically homogeneous evolution is expected to operate primarily (if not exclusively) at low metallicity and to preferentially produce high-mass ($\gtrsim 25,\msun$) \acrshort{bh}. In addition, independent of the formation channel, more massive \acrshortpl{bh} are expected to form more efficiently in metal-poor environments due to the strong dependence of wind mass loss rates of their stellar progenitors on metallicity \citep[e.g.,][]{2009MNRAS.395L..71M,2010ApJ...714.1217B,2015MNRAS.451.4086S,2020MNRAS.498..495D,2021MNRAS.504..146V,2024ApJ...964L..23R,2025arXiv250717052M}.
Therefore, the redshift evolution of the \acrshort{bh} merger rate is highly uncertain, but expected to vary across the mass spectrum, and the role of metallicity in shaping this evolution likely differs between our low- and high-mass bins.

\begin{figure*}
    \centering
    \includegraphics[width=\textwidth]{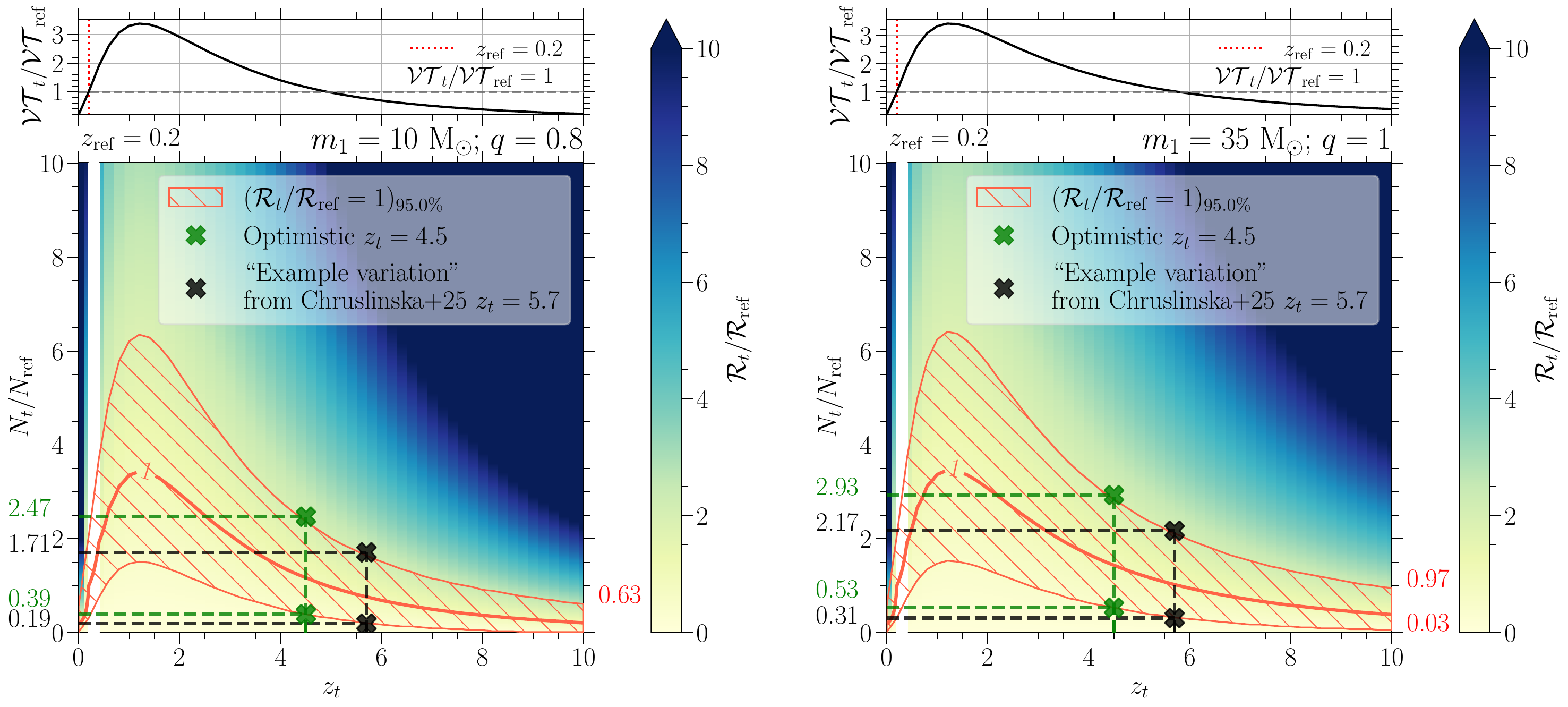}
    \includegraphics[width=\textwidth]{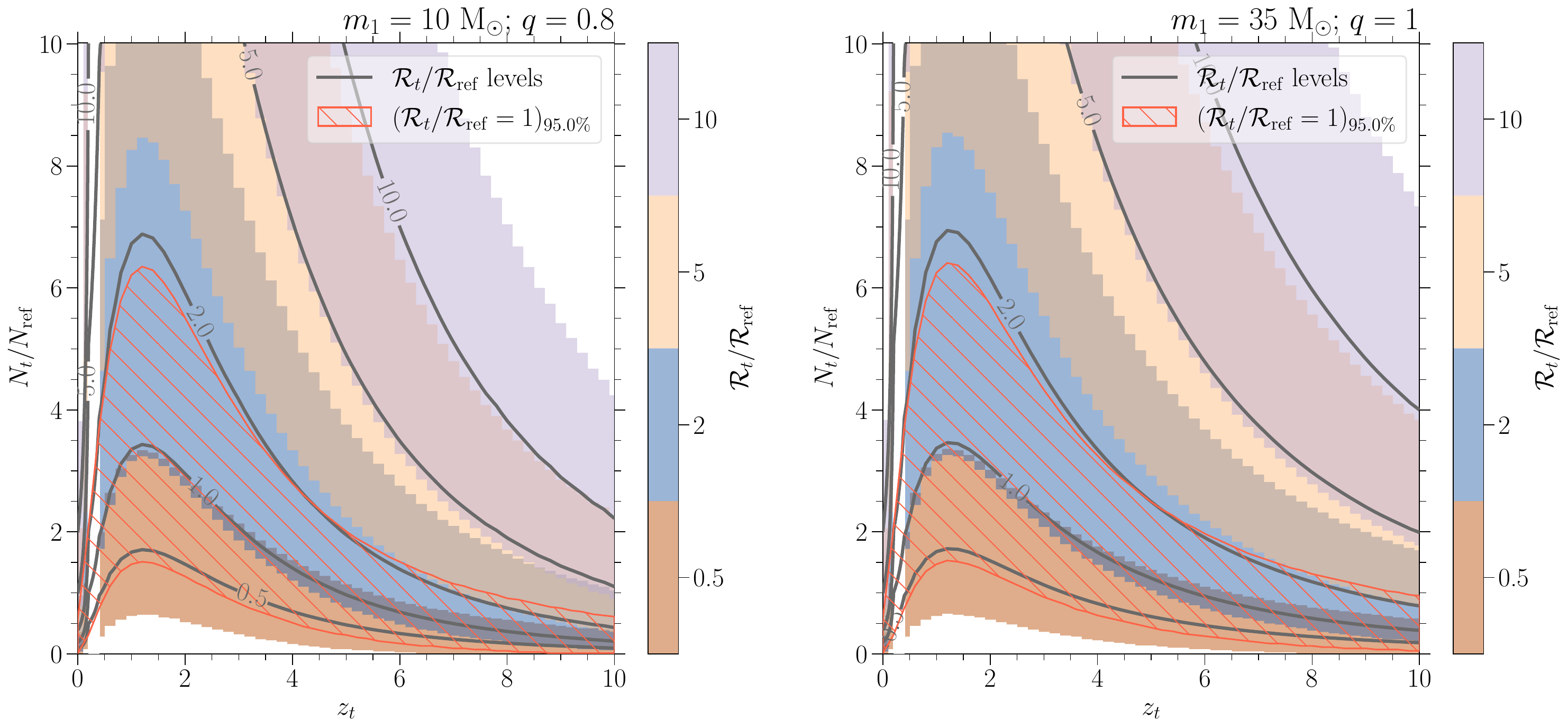}
    \caption{Inferred merger rate ratio across target redshift and event statistics. The top panels show the median of the inferred \(\Rate_t / \Rate_\reflb\) distribution (color scale), obtained from~\cref{eq:Rate_ratio_posterior} as a function of the target redshift (horizontal axis), for a fixed reference redshift, \(\zref = 0.2\). The secondary panels at the top show the corresponding evolution of the sensitivity-volume ratio, \(\VT_t / \VT_\reflb\). 
The bottom panels show 95\% confidence regions for selected values of \(\Rate_t / \Rate_\reflb\). These regions were constructed by selecting bins for which the 95\% confidence interval of the inferred \(\Rate_t / \Rate_\reflb\) distribution contains the values indicated on the color bar. 
The left and right panels correspond to fixed-mass bins: \(m_1 \in [7, 13]\,\msun\) with \(q = 0.8\), and \(m_1 \in [32, 38]\,\msun\) with \(q = 1\), respectively. Redshift bins have width \(\Delta z = 0.2\), the observation time is set to \(\Tobs = 1\,\yr\), and the number of events in the reference bin is fixed to \(N_\reflb = 10\). The hatched region marks the confidence interval for \(\Rate_t / \Rate_\reflb = 1\). 
Green and black crosses along the boundary of the hatched region indicate the minimum detection statistics required to conclude \(\Rate_t / \Rate_\reflb \neq 1\) at 95\% credibility for a given \(z_t\). The same information is reported by red labels on the right vertical axis for \(z_t = 10\). Bins with \(z_t = \zref\) are masked, as \(\Rate_t / \Rate_\reflb = 1\) by construction.\label{fig:N_zt_map}}
\end{figure*}

\begin{figure*}
    \centering
    \includegraphics[width=\textwidth]{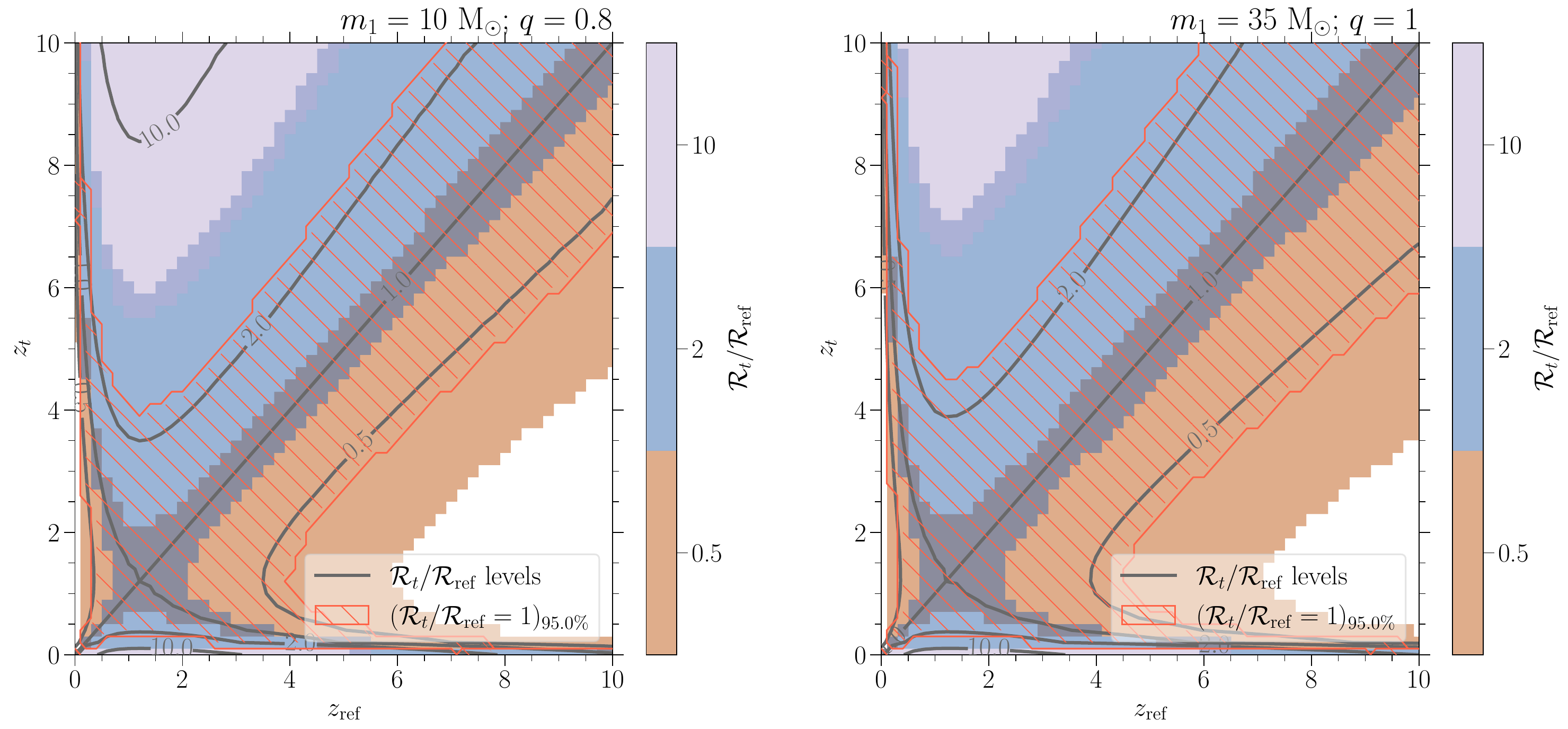}
    \caption{95\% confidence regions for the merger rate density ratio for selected values of \(\Rate_t / \Rate_\reflb\), assuming \(N_t / N_\reflb = 1\). See~\cref{fig:N_zt_map} for details on the construction of the confidence regions and the adopted binning. We assume a fixed observation time of \(\Tobs = 1\,\yr\) and \(N_\reflb = 10\). This choice is appropriate for \(\zref = 0.2\) \citep{2026JCAP...03..081A} and is extended to all values of \(\zref\) shown on the horizontal axis for simplicity. A more accurate treatment, accounting for the variation in \(N_\reflb\) with \(\zref\), would result in confidence regions with different widths.}\label{fig:zt_zref_map}
\end{figure*}

Motivated by these considerations, we defined low- and high-mass bins centered at $10\,\msun$ and $35\,\msun$, respectively, based on the primary mass of the binary. To present the merger rate density ratio results, we fixed the secondary \acrshort{bh} mass by assuming constant mass-ratio ($q$) systems. We chose $q$ based on results by~\cite{2025arXiv250818083T}: \acrshortpl{bh} with masses of $\sim 10\,\msun$ preferentially merge with lighter companions ($q \sim 0.8$), while those with masses of $\sim 35\,\msun$ tend to merge with more equal-mass partners ($q \sim 1$). 
This simplified setup is intended to illustrate the key features of our method and its ability to enable rapid, informative comparisons between merger rates at different redshifts.

\subsection{Constraining the merger-rate density ratio}\label{sec:Rate_ratio_inference}

We now present the results from our merger rate density ratio estimator.
We fixed the reference redshift to $z_\reflb = 0.2$, which allowed us to anchor our estimates to the mass-specific merger rate measured by LVK and avoid extrapolating the uncertain redshift evolution of binary \acrshort{bh} population properties. The expected number of detections in the reference bin, $N_\reflb$, over a given observing time, $\Tobs$, then follows from the assumed sensitivity of next-generation detectors.
The results discussed in this section assume $\Tobs = 1\, \yr$, and the corresponding number of events, $N_\reflb$, associated with the reference bin is fixed to $10$ (estimated from Fig.~3.5 by \citealt{2026JCAP...03..081A} at $\zref = 0.2$). 
The confidence interval of the merger rate density ratio depends on both the number of detections falling within the considered bins and their associated \acrlong{vt}. 
To show the behavior of~\cref{eq:Rate_ratio_posterior} in different scenarios, we produced two sets of plots: we show both the $N_t / N_\reflb$ versus\ $z_t$ plane (\cref{fig:N_zt_map}) assuming a fixed $\zref$; and the $z_t$ versus\ $\zref$ plane (\cref{fig:zt_zref_map}) assuming a fixed $N_t / N_\reflb$.

\Cref{fig:N_zt_map} is divided into two panels illustrating complementary aspects of~\cref{eq:Rate_ratio_posterior}. In the upper panels, the color map shows the median of the inferred \(\Rate_t / \Rate_\reflb\) distribution. The top secondary panels show the corresponding \acrlong{vt} ratios as a function of redshift, providing a direct mapping between the ratio of detected events and the merger rate ratio through~\cref{eq:Rate_ratio}. For visualization purposes, values of \(\Rate_t / \Rate_\reflb > 10\) are saturated at the edge of the color map; within the explored parameter space, this affects only the upper right corner of the plots, where \(\Rate_t / \Rate_\reflb \sim 10^2\). 
In the lower panels, each bin is colored according to whether the 95\% confidence interval of the corresponding \(\Rate_t / \Rate_\reflb\) distribution -- computed using~\cref{eq:Rate_ratio_posterior} -- includes selected reference values indicated on the color bar. This construction yields a set of colored ``confidence regions'' centered on specific values of \(\Rate_t / \Rate_\reflb\). In all panels, the hatched region corresponds to the confidence interval for \(\Rate_t / \Rate_\reflb = 1\). 
Because these regions overlap, the key information lies in identifying where the merger rate can be distinguished from a given reference value. For example, outside the hatched region one can conclude that \(\Rate_t / \Rate_\reflb \neq 1\) at the 95\% confidence level. 

\begin{table}
    \caption{Astrophysical merger rate variation.}
    \centering
    \renewcommand{\arraystretch}{1.4}
    \begin{tabular}{lcc}
        \toprule
        & $10\, \msun$ & $35\, \msun$ \\
        \midrule
        \multirow{2}{*}{$z_t = 4.5$} & $2.03^{+2.36}_{-1.02}$ & $1.98^{+2.25}_{-0.97}$ \\
        \addlinespace
        & $0.33^{+0.64}_{-0.24}$ & $0.37^{+0.63}_{-0.25}$ \\
        \midrule
        \multirow{2}{*}{$z_t = 5.7$} & $2.15^{+2.68}_{-1.14}$ & $2.05^{+2.44}_{-1.05}$ \\
        \addlinespace
        & $0.26^{+0.69}_{-0.22}$ & $0.31^{+0.66}_{-0.23}$ \\
        \midrule
        \multirow{2}{*}{$z_t = 10$} & $2.90^{+4.70}_{-1.87}$ & $2.46^{+3.50}_{-1.45}$ \\
        \addlinespace
        & N.A. & $0.13^{+0.81}_{-0.12}$ \\
        \bottomrule
    \end{tabular}
    \tablefoot{Astrophysical merger rate variation, $\Rate_t / \Rate_\reflb$, at the marked points of~\cref{fig:N_zt_map} for target redshifts $z_t = 4.5$, $5.7$, and $10$, in the two considered mass bins. For each target redshift, the upper and lower values refer to the top and bottom markers in~\cref{fig:N_zt_map}, respectively. The corresponding distributions are shown in~\cref{fig:Rratio_pdf}.}\label{tab:Rt_Rref}
\end{table}

\Cref{fig:zt_zref_map} shows the same confidence regions as the bottom panels of~\cref{fig:N_zt_map}, but for a fixed value of \(N_t / N_\reflb = 1\). While our primary interest lies in the low-\(\zref\) regime, this representation provides a more general view of merger rate density variations between arbitrary bins. 
For simplicity, we fix \(N_\reflb = 10\) for all values of \(\zref\). In practice, however, \(N_\reflb\) is expected to vary with \(\zref\) due to the redshift evolution of the astrophysical rate: for fixed bin widths and observation time, \(N_\reflb\) is larger where \(\Rate_\reflb\) is higher. This would, in turn, affect the confidence regions, as both \(N_\reflb\) and \(N_t\) determine the width of the posterior distributions obtained from~\cref{eq:Rate_ratio_posterior}.

In the top panels of~\cref{fig:N_zt_map}, we highlight explicitly the detection count ratios (green and black crosses) needed to infer a minimum merger rate variation, assuming $N_\reflb = 10$. In the $10\,\msun$ case, $N_t$ must differ from $N_\reflb$ by a factor of $2.5$ ($1.7$) for $z_t = 4.5$ ($z_t = 5.7$). In the $35\,\msun$ case, the corresponding values of $N_t / N_\reflb$ are $2.9$ and $2.2$, respectively. The implied variation in the astrophysical rate, $\Rate_t / \Rate_\reflb$, is reported in~\cref{tab:Rt_Rref}. 
The bottom panels of~\cref{fig:N_zt_map} illustrate the overlap of the confidence regions for selected values of $\Rate_t / \Rate_\reflb$. Reducing the width of these regions -- and thus better constraining the rate variation between $z_t$ and $\zref$ -- requires higher detection statistics. In the $N_t / N_\reflb$ vs.\ $z_t$ plane, the width of the confidence regions is primarily controlled by $N_\reflb$. For example, for $N_t / N_\reflb = 1$,~\cref{fig:LRT_vs_Bayes} shows that the confidence interval of the $\Rate_t / \Rate_\reflb$ distribution shrinks by a factor of $3.57$ when $N_\reflb$ increases from $10$ to $100$. This directly translates into tighter confidence regions in~\cref{fig:N_zt_map}.

\begin{figure}
    \centering
    \includegraphics[width=\columnwidth]{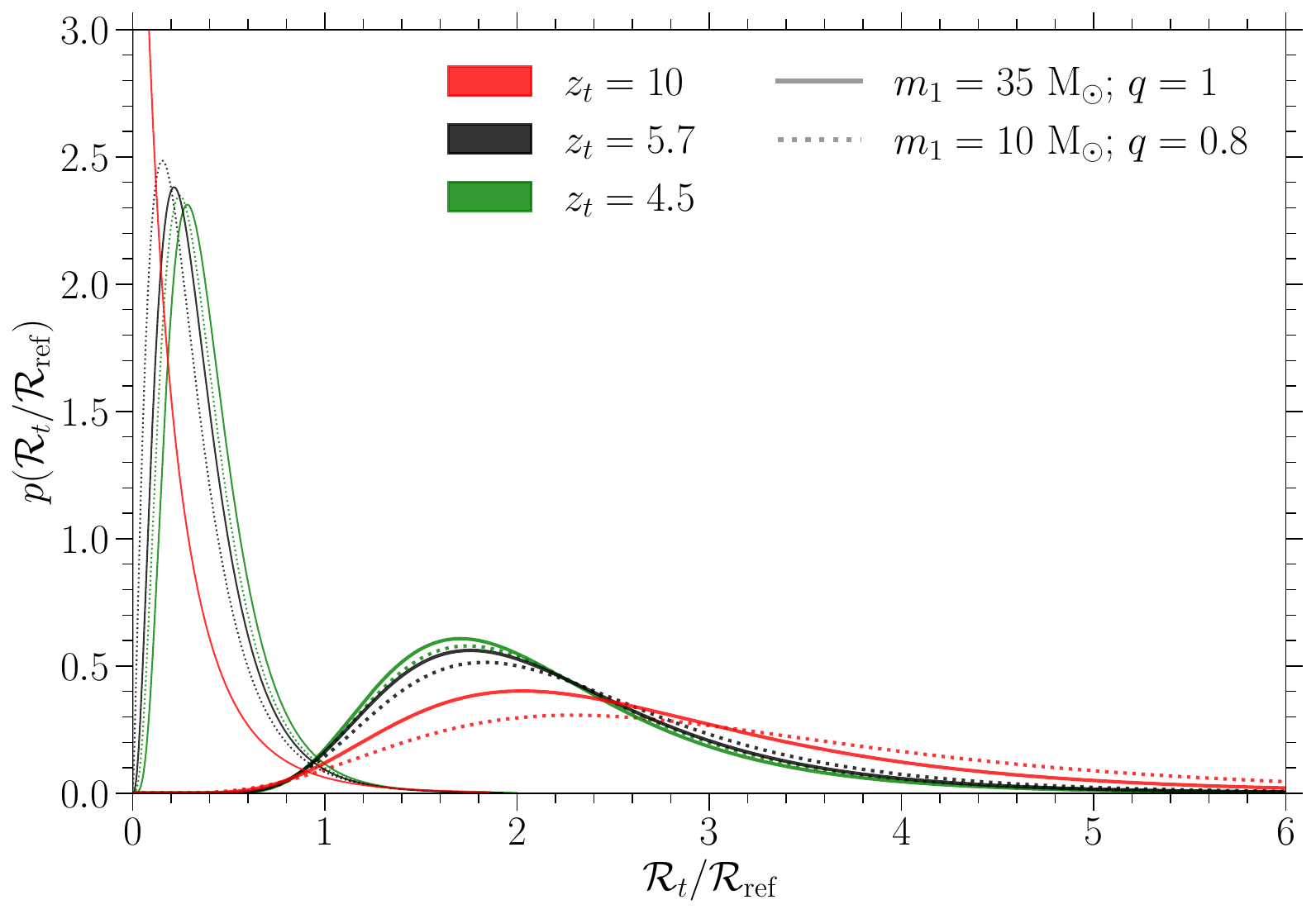}
    \caption{Distributions of $\Rate_t / \Rate_\reflb$ inferred using~\cref{eq:Rate_ratio_posterior} for the points marked in~\cref{fig:N_zt_map}. Each distribution corresponds to a specific combination of target redshift (indicated by color) and primary \acrshort{bh} mass (solid or dashed lines). Thick and thin lines denote the upper and lower boundaries of the hatched region in~\cref{fig:N_zt_map}, respectively. The $95\%$ confidence intervals for each distribution are listed in~\cref{tab:Rt_Rref}.}\label{fig:Rratio_pdf}
\end{figure}

For completeness,~\cref{fig:Rratio_pdf} provides a complementary visualization of the $\Rate_t / \Rate_\reflb$ distributions corresponding to the points marked in~\cref{fig:N_zt_map}, including the case $z_t = 10$. We note that these results are not corrected for the integer nature of the number of detections (see Appendix~\ref{sec:apx_LRT_vs_Bayes}). This approximation primarily affects $\Rate_t / \Rate_\reflb$ when the expected number of detections is below unity (which is the case for the thin solid red line in~\cref{fig:Rratio_pdf}).

Finally, we can use our results to quantify the relative evolution of the merger rate with redshift across the two mass bins. We define
\begin{align}\label{eq:D_ratio_ratio}
   \mathcal{D}\ =&\
   \dfrac{\Rate_t (10\, \msun) / \Rate_t (35\, \msun)}{\Rate_\reflb (10\, \msun) / \Rate_\reflb (35\, \msun)}
\end{align}
so that $\mathcal{D} \neq 1$ would indicate a differential evolution of the merger rate between low- and high-mass binaries. We note that if the detection statistics in the two bins are at the minimum thresholds required to claim a merger rate variation between target and reference redshifts, then $\mathcal{D} \sim 1$ for all considered $z_t$. The result is only weakly dependent on $N_\reflb$, which mainly affects the symmetry and kurtosis of the $\mathcal{D}$ distribution. Specifically, we obtain $\mathcal{D} = 1.0^{+1.8}_{-0.7}$, $1.0^{+2.0}_{-0.7}$, and $1.2^{+3.2}_{-0.9}$ for $z_t = 4.5$, $5.7$, and $10$, respectively.

\section{Discussion}
There is no unique target redshift at which low-metallicity environments can be isolated.
Even with our modest requirement of a low metallicity of 10\% of the solar iron abundance,  current models span a wide range, from \(z_t \sim 4\) to \(z_t > 10\). If the ``pessimistic'' scenario (fast \(\Fe\) enrichment) holds, the relevant redshift window shifts to such high values that, despite the excellent sensitivity of future detectors, uncertainties in source parameters may hinder the identification of specific subpopulations. 
The ability of next-generation detectors to probe stellar physics will therefore also rely on external constraints on the \acrshort{sfh}. While the reach of ET+CE is sufficient to observe massive binary \acrshortpl{bh} out to the cosmic dawn, \cref{fig:N_zt_map,fig:zt_zref_map} shows that it may remain challenging to determine conclusively whether \acrshort{bh} formation physics differs from that in the local Universe, depending on the target redshift and the available detection statistics. While a detailed comparison of different ET geometries is beyond the scope of this work, a 2L configuration with \(\geq 15\ \mathrm{km}\) arms is expected to yield higher detection rates and improved population constraints \citep{2023JCAP...07..068B}. Because our methodology relies on the relative ratio of merger-rate densities across mass and redshift bins, absolute differences in detector sensitivity are largely mitigated by the \(\VT_\reflb / \VT_t\) term of~\cref{eq:Rate_ratio_posterior}, leaving our qualitative conclusions unchanged. Quantitatively, the higher number of expected detections, \(N_\reflb\), provided by a more sensitive 2L configuration would narrow the confidence intervals of the inferred merger-rate ratios (as shown in Appendix~\ref{sec:apx_LRT_vs_Bayes}), thereby reducing the minimum observational time required to establish a significant rate variation.
Finally, a deviation from unity in \(\Rate_t / \Rate_\reflb\) should not be interpreted as a pure metallicity signature, but rather as the combined effect of formation efficiency, \acrlong{dtd}, and the \acrshort{sfh}. Moreover, systems formed through dynamical channels -- such as in dense stellar clusters or active galactic nucleus disks -- are governed by dynamical assembly mechanisms that can significantly alter their delay times. These processes enable such binaries to merge at \(z_t\), consequently polluting the high-redshift sample and obscuring the metallicity signature of the isolated population.
Stellar-evolution models predict a redshift-dependent merger rate density that varies with \acrshort{bh} mass \citep[e.g.,][]{2022ApJ...931...17V,2024ApJ...976...23B,2026ApJ...997..231S}, reflecting the underlying formation physics.

\begin{figure}
    \centering
    \includegraphics[width=\columnwidth]{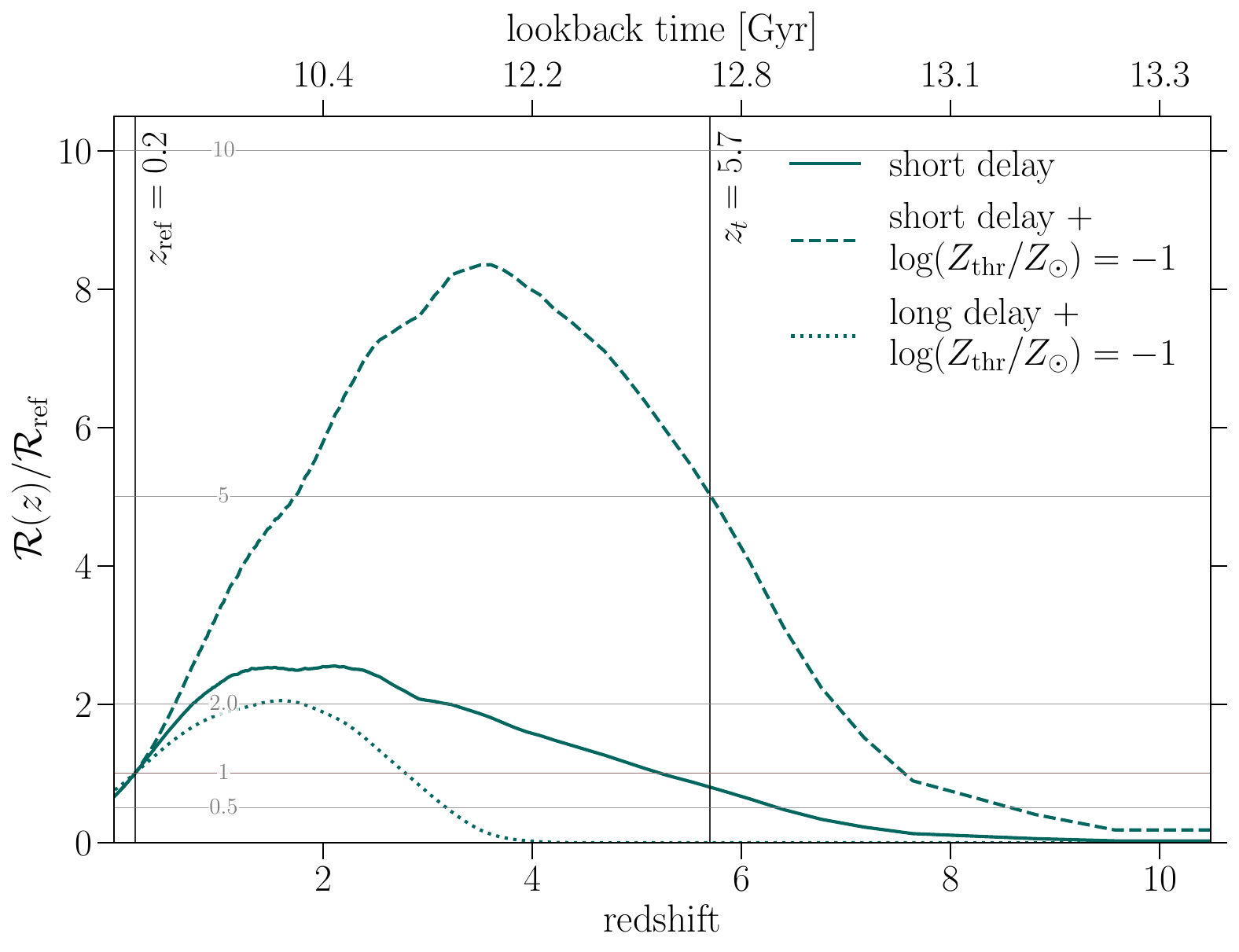}
    \caption{Examples of \Acrshort{bh} merger rate densities computed convolving the metallicity-dependent cosmic \acrshort{sfh} with a metallicity-dependent formation efficiency and a delay-time distribution (see text). Curves are normalized to unity at a reference redshift, $z_\reflb=0.2$.}\label{fig:rate_examples}
\end{figure}

The results presented in Sec.~\ref{sec:Rate_ratio_inference} are deliberately agnostic about the underlying astrophysical population model: our aim is to propose a test of such models rather than to assume one. 
For concreteness, in~\cref{fig:rate_examples} we show the merger-rate-density ratio as a function of redshift $\Rate(z)/\Rate(z_\reflb)$ for the illustrative examples from \citet{2025arXiv251115782C};
details are reported in Appendix~\ref{sec:apx_rate_toy_model} below.
These examples were obtained by convolving the metallicity-dependent cosmic \acrshort{sfh} with either metallicity-dependent or metallicity-independent formation efficiencies, $\eta$, and a conventional delay-time distribution, $\propto t^{-1}$. The minimum delay time was varied between a ``short-delay'' case, $\tau_{\mathrm{min}}=10\,\mathrm{Myr}$, and a ``long-delay'' case, $\tau_{\mathrm{min}}=1\,\mathrm{Gyr}$. 
The formation efficiency (number of merging BHs formed per unit of stellar mass formed) is described with a logistic function,
\begin{equation}
\eta(Z)
=
\eta_\mathrm{high}
+
\frac{\eta_\mathrm{low}-\eta_\mathrm{high}}
{1+\exp\left[-k\left(Z-Z_\mathrm{thr}\right)\right]},
\label{eq:metallicity_efficiency}
\end{equation}
where $Z_\mathrm{thr}$ is the threshold metallicity at which the efficiency transitions from the low-metallicity plateau, $\eta_\mathrm{low}$, to the high-metallicity value, $\eta_\mathrm{high}$. We used $k=-10$ and assumed $\eta_\mathrm{high} = 10^{-2}\eta_\mathrm{low}$ so that the formation efficiency drops by two orders of magnitude above the threshold \citep{2025arXiv251115782C}.

We show the examples for the iron-dependent cosmic \acrshort{sfh} model that yields a moderate target redshift $z_t=5.7$ to select metal-poor progenitors (corresponding to the example variation by \citealt{2025arXiv251115782C}).
Although the formation efficiency is generally enhanced at low metallicity, most models predict either a weaker metallicity dependence or a higher effective metallicity threshold than $\log(Z_\mathrm{thr}/Z_{\odot})=-1$, as assumed in the examples shown in~\cref{fig:rate_examples}.
Exceptions include channels focused specifically on extremely metal-poor or
Population~III progenitors, which arise from rare star-forming environments and are beyond the scope of this discussion.

The delay-time prescriptions considered here also bracket a range of expected merger timescales: stable mass-transfer channel can produce characteristic delays on the order of $\sim 1\,\mathrm{Gyr}$, common-envelope channel can yield much shorter delays, and other
channels may populate the intermediate regime. We point out that delay times of $\sim 1\,\mathrm{Gyr}$ are comparable to the age of the Universe at $z_t$, and therefore
yield $\Rate(z_t)/\Rate(z_\reflb) \approx 0$ regardless of $\eta$, as is illustrated by the dotted line in~\cref{fig:rate_examples}. For short delay times, instead, $\Rate(z_t)/\Rate(z_\reflb) \gtrsim 1$, with larger values if $\eta$ is enhanced below a low metallicity threshold (solid and dashed lines in~\cref{fig:rate_examples}). 
The realistic \acrshort{bh} merger population is likely to arise
from a mixture of channels, with different delay times and formation efficiencies, and with different relative contributions in the low- and high-mass bins (see Sec.~\ref{sec:mass_features}). Therefore, distinguishing between most current formation models, and identifying differences between the mass bins at $z_t$, would require one to constrain $\Rate(z_t)/\Rate(z_\reflb)$ to within a factor of a few.

In Sec.~\ref{sec:Rate_ratio_inference}, we showed, using the estimator, $\mathcal{D}$, of~\cref{eq:D_ratio_ratio}, that achieving the detection statistics required to establish a minimum merger rate variation within individual mass bins is not sufficient to demonstrate a differential evolution across masses.
In this context, our framework provides a fast way to assess the detection requirements needed to test stellar evolution predictions for the binary \acrshort{bh} merger rate density. While matching the relative number of detections across masses and redshifts to theoretical expectations can be formally reduced to applications of~\cref{eq:Rate_ratio}, we argue that such ratios should be interpreted in light of the minimum detection thresholds identified in~\cref{fig:N_zt_map}, i.e., the requirements needed to establish a rate evolution between given redshifts in the first place.
Furthermore, we find that achieving an uncertainty on $\mathcal{D}$ at the level of $\sim 50\%$ ($\sim 10\%$), based on its 95\% credible interval, requires $\sim 200$ ($\sim 4000$) detections in the reference bin, $N_\reflb$.

The analysis presented here focuses on mass bins centered at $10\,\msun$ and $35\,\msun$, motivated by the features identified with current \acrshort{lvk} data.
Future applications could target different mass ranges. The choice of reference redshift is also important, as $N_\reflb$ sets the width of the confidence regions in~\cref{fig:N_zt_map,fig:zt_zref_map}.
Our framework currently compares a single pair of bins, but it can be extended to incorporate multiple locations in parameter space.
An additional observational challenge for our methodology is the accuracy with which the redshift of binary \acrshort{bh} mergers can be inferred \citep{2022ApJ...931L..12N,2023PhRvD.107j1302M}. For \acrshort{gw} sources without electromagnetic counterparts, the redshift is not measured directly but is derived from the luminosity distance assuming a standard cosmological model. For a network of ET and 2 CE, about \(90\%\) of the \acrshort{bh} population will be detected with \(\Delta d_L / d_L \lesssim 0.2\), while at \(z \gtrsim 5\) only about \(7\%\) will have \(\Delta d_L / d_L \lesssim 0.02\) \citep{2022ApJ...941..208I}. Therefore, at the target redshifts explored in this work (\(z_t \gtrsim 4\)), the uncertainties on the luminosity distance can translate into broad posterior distributions for the inferred redshift. The consequent cross-contamination between nearby redshift bins may effectively blur the boundaries of the selected subpopulations.

Our framework relies on a simplified counting approach that neglects individual parameter-estimation uncertainties. Fully incorporating these effects requires computationally intensive hierarchical analyses \citep{2019MNRAS.486.1086M,2022hgwa.bookE..45V} that can be adapted to select events in a given mass or redshift range \citep[e.g.,][]{2025CQGra..42v5008K}. Nevertheless, our approximation is adequate for rapid forecasting studies aimed at building the science case and informing the design of next-generation interferometer networks \citep{2021arXiv210909882E,2026JCAP...03..081A}.

\section{Summary}
Next-generation \acrshort{gw} detectors, such as ET and CE, will extend the observational horizon for stellar-mass \acrshort{bh} mergers far beyond current limits, revealing previously unexplored \acrshort{bh} populations. This increased reach opens new opportunities for stellar astrophysics: by selecting sources at specific redshifts, one can probe remnants originating from progenitors formed in early cosmic environments, thereby isolating the impact of low metallicity.

In this work, we present a statistical framework to quantify the precision required to compare binary \acrshort{bh} populations at a target redshift, $z_t$, with the one observed at $\zref=0.2$. Specifically:
\begin{itemize}
    \item We developed an approach based on Bayesian inference (alongside a \acrlong{lrt}) to estimate the merger rate density ratio between two mass--redshift bins, accounting for selection effects through the differential space-time volume.
    \item To illustrate the method, we focused on binary \acrshort{bh} subpopulations with primary masses of $10\, \msun$ and $35\, \msun$, corresponding to overdensities observed in the \acrshort{bh} mass spectrum with \acrshort{lvk} data.
    \item Using updated models of the metallicity-dependent cosmic \acrshort{sfh}, we showed that ensuring the majority of progenitors form in low-metallicity environments ($Z < 0.1 Z_\odot$) requires target redshifts of $z_t > 4$, potentially extending to $z_t > 10$ depending on model uncertainties.
\end{itemize}
While physically motivated choices of the target redshift can help isolate \acrshort{bh} progenitors formed under specific environmental conditions -- thereby enhancing the power of \acrshortpl{gw} as probes of stellar evolution -- constraining merger-rate variations with high precision may remain challenging. The differences between the population properties observed at target and reference redshift may not be due to metallicity alone, but may be due to a combination of factors (e.g., a contribution from dynamically formed mergers in dense stellar environments). This complicates the interpretation. Additional selection criteria may be required to distinguish between subpopulations formed in different channels.
Within this framework, an accurate characterization of the local \acrshort{bh} population will continue to play a central role in determining $\Rate_t / \Rate_\reflb$. 
The simplicity of the approach makes it a reusable tool for rapid bin-to-bin comparisons in future population studies, as well as for testing stellar evolution model predictions against observations from next-generation \acrshort{gw} detectors.

\begin{acknowledgements}

We thank Ssohrab Borhanian for discussions.
F.LdP. and D.G. are supported by 
ERC Starting Grant No.~945155--GWmining, 
Cariplo Foundation Grant No.~2021-0555, 
Italian-French University (UIF/UFI) Grant No.~2025-C3-386,
and
MUR Grant ``Progetto Dipartimenti di Eccellenza 2023-2027'' (BiCoQ).
F.LdP. is supported by an Erasmus+ scholarship.
D.G. is supported by 
MSCA Fellowship No.~101149270--ProtoBH, 
MUR Young Researchers Grant No. SOE2024-0000125, 
and the INFN TEONGRAV initiative.
We gratefully acknowledge the Max Planck Institute for Astrophysics for hosting F.LdP. during part of this research project.
Computational work was performed
at CINECA with allocations through INFN and the University of Milano-Bicocca.
\end{acknowledgements}

\bibliographystyle{aa_edited}
\bibliography{targetredshift}

\begin{appendix}

\section{Distributions and properties}\label{sec:apx_bayes}
Below, we list the definitions of some probability density functions used in this work.

\begin{itemize}
\item
Poisson distribution with mean $\lambda$:
\begin{equation}\label{eq:poisson_dist}
    \mathrm{Poisson}(k; \lambda) = \dfrac{\lambda^k \exp(-\lambda)}{k!}\,.
\end{equation}
\item
Gamma distribution with shape parameter $\alpha$ and rate parameter $\beta$:
\begin{equation}\label{eq:gamma_dist}
    \mathrm{Gamma}(x; \alpha, \beta) = \dfrac{\beta^\alpha}{\Gamma(\alpha)} x^{\alpha-1} \exp(-\beta x)\,,
\end{equation}
where $\Gamma$ is the gamma function.
\item
Generalized beta prime distribution:
\begin{equation}\label{eq:betaprime_dist}
    \beta'(x; \alpha, \beta, p, q) = \dfrac{p \qty(\dfrac{x}{q})^{\alpha p - 1} \qty[1 + \qty(\dfrac{x}{q})^p]^{-\alpha -\beta}}{q B(\alpha, \beta)}\,,
\end{equation}
where $B$ is the beta function. In particular, if $X \sim \beta'(\alpha, \beta, p, q)$ and $k > 0$, then
\begin{equation}\label{eq:betaprime_prop1}
    kX \sim \beta'(\alpha, \beta, p, kq)\,.
\end{equation}
\end{itemize}

\section{Likelihood-ratio test}\label{sec:apx_lrt}
The \acrlong{lrt} requires computing the test statistic
\begin{equation}
\lambdaLR = -2 \qty[\ln \mathcal{L}\qty(\hat{\theta}_0) - \ln \mathcal{L}\qty(\hat{\theta}_1)]\,,
\end{equation}
where $\hat{\theta}_0$, $\hat{\theta}_1$ are the \acrlong{mle} for the likelihood parameters under $\mathcal{H}_0$ and $\mathcal{H}_1$. We defined the likelihood as $\mathcal{L}\qty(\lambda_t, \lambda_\reflb) = \mathrm{Poisson}\qty(\lambda_t) \times \mathrm{Poisson}\qty(\lambda_\reflb)$ and reparametrized it using $\lambda_\reflb$ and $\alpha = \lambda_t / \lambda_\reflb$. One has
\begin{align}
    \ln \mathcal{L}\qty(\alpha, \lambda_\reflb) &= 
    N_t\ln(\alpha) + N_t\ln(\lambda_\reflb) - \alpha\lambda_\reflb + N_\reflb\ln(\lambda_\reflb) 
    \notag \\
    &\quad - \lambda_\reflb - \ln(N_t!N_\reflb!)\,.
\end{align}
Maximization over both $\alpha$ and $\lambda_\reflb$ leads to $\alpha = {N_t}/{N_\reflb}$ and $\lambda_\reflb = N_\reflb$ and thus
\begin{equation}
    \mathcal{L}\qty(\hat{\theta}_1) = \dfrac{N_t^{N_t}}{N_t!} \dfrac{N_\reflb^{N_\reflb}}{N_\reflb!} \exp(-N_t-N_\reflb)\,.
\end{equation}
Maximization over $\lambda_\reflb$ imposing $\alpha = \alpha_0$ yields $\tilde{\lambda}_\reflb = ({N_t + N_\reflb})/({1 + \alpha_0})$, from which we obtained
\begin{equation}\mathcal{L}\qty(\hat{\theta}_0) = \dfrac{\qty(\alpha_0 \tilde{\lambda}_\reflb)^{N_t}}{N_t!} \exp\qty(-\alpha_0\tilde{\lambda}_\reflb) \dfrac{\tilde{\lambda}_\reflb^{N_\reflb}}{N_\reflb!} \exp\qty(-\tilde{\lambda}_\reflb)\,.
\end{equation}
Using the definitions above, the test statistic is given by
\begin{equation}
\lambdaLR = 2 N_\reflb \qty[\alpha\ln\qty(\dfrac{\alpha}{\alpha_0}) - \qty(1 + \alpha) \ln\qty(\dfrac{1 + \alpha}{1 + \alpha_0})]\,,
\end{equation}
which is~\cref{eq:lambda_lr_log}.
The equation above can be rewritten as
\begin{equation}\label{eq:lambda_lr_exp}
    \qty[\exp\qty(\dfrac{\lambdaLR}{2 N_\reflb}) \qty(1 + \alpha)\qty(1 + \dfrac{1}{\alpha})^\alpha \alpha_0^\alpha  ]^{\dfrac{1}{1 + \alpha}} - \alpha_0 - 1 = 0\,,
\end{equation}
which is preferable for root-finding algorithms.

\section{Likelihood-ratio test vs. Bayesian inference}\label{sec:apx_LRT_vs_Bayes}

\begin{figure}
    \centering
    \includegraphics[width=\columnwidth]{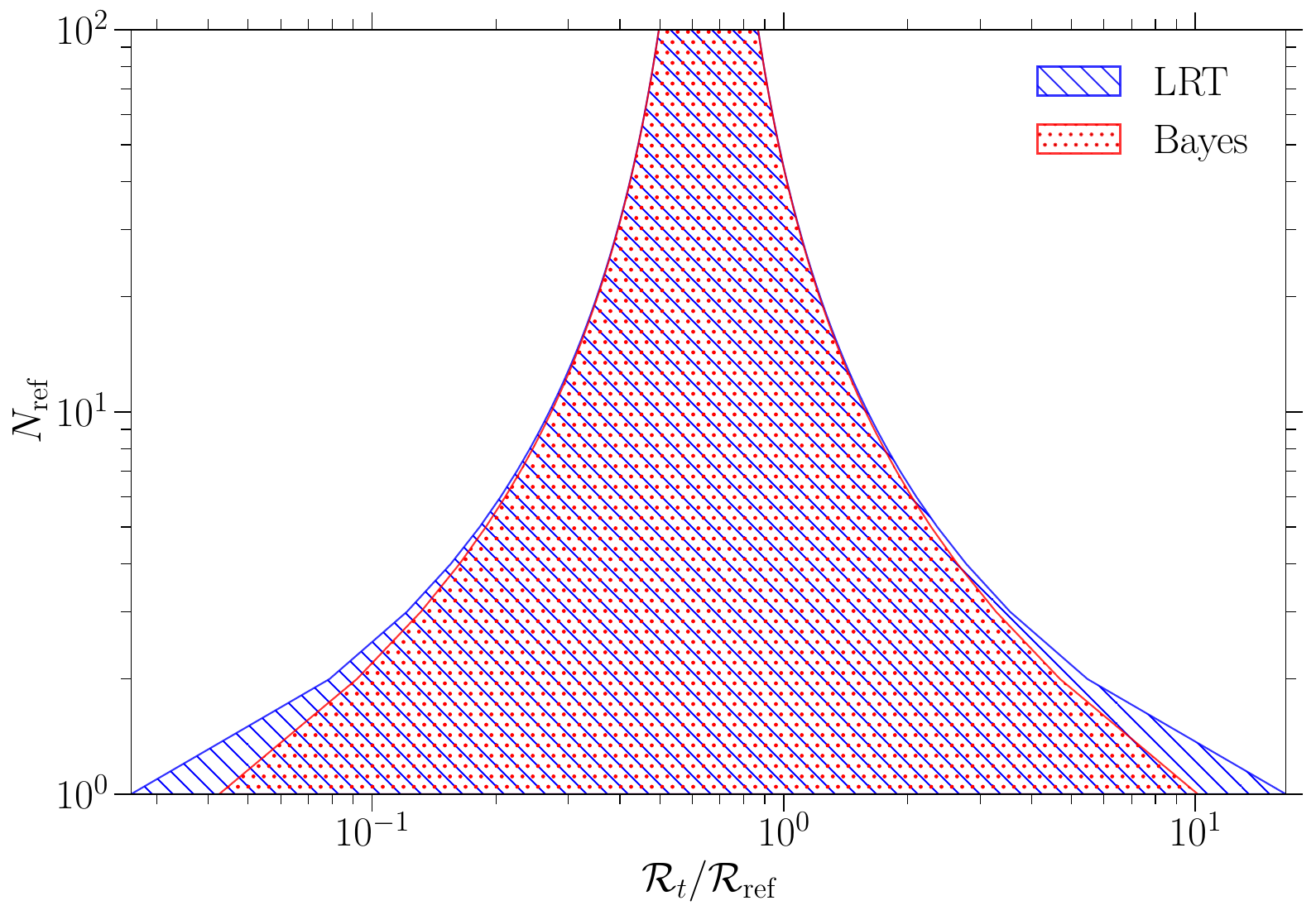}
    \caption{Illustrative example of the confidence intervals obtained with both the \acrlong{lrt} and~\cref{eq:Rate_ratio_posterior} assuming $N_t / N_\reflb = 1$, for various values of $N_\reflb$. 
    The comparison is done for the $m_1 = 10\, \msun$ mass bin and redshifts $z_t = 5.7$ and $\zref = 0.2$. 
    Bin's widths are the same as in~\cref{fig:N_zt_map}.
    For a given $N_\reflb$, the boundaries of the shaded areas mark the extremes of the $95\%$ confidence interval for $\Rate_t / \Rate_\reflb$.
    For instance, for $N_\reflb = 10$ ($100$) this corresponds to $\Rate_t / \Rate_\reflb \in [0.27, 1.59]$ ($[0.50, 0.87]$).}\label{fig:LRT_vs_Bayes}
\end{figure}

{In~\cref{fig:LRT_vs_Bayes}} we compare the confidence intervals obtained with the two methods introduced in this paper: the \acrlong{lrt} and the Bayesian approach {of~\cref{eq:Rate_ratio_posterior}}. For simplicity, we assumed a measured event ratio $N_t / N_\reflb = 1$. The figure shows that the two approaches yield nearly identical confidence intervals for sufficiently large values of $N_\reflb$. 
In the example shown, noticeable differences are present only when $N_\reflb$ is of order unity, where the asymptotic regime required by Wilks' theorem (see Sec.~\ref{sec:lrt}) is not satisfied. In this regime, the Bayesian approach is preferable, as it does not rely on asymptotic assumptions and depends only on the choice of prior for the Poisson rate parameter (see Sec.~\ref{sec:bayes}) associated with each bin.

The results in this paper are presented using the Bayesian approach. We note that~\cref{eq:Rate_ratio_posterior} is formally defined for integer values of $N_t$ and $N_\reflb$, which we have extended to real numbers. We have verified that this approximation has a negligible impact on the results presented here.

\section{Delay-time convolution}\label{sec:apx_rate_toy_model}

Following Sec.~7.4 of~\cite{2025arXiv251115782C}, we considered a few examples to discuss the possible \acrshort{bh} merger rate density evolution with redshift.
The rate is calculated by convolving the metallicity-dependent cosmic \acrshort{sfh} with a metallicity-dependent formation efficiency and a delay-time distribution:
\begin{equation}
\Rate_{}(z)
=
\int_{0}^{z(t_0)}
\int
f_{\mathrm{SFR}}(Z_{\Fe/\mathrm{H}},t')
\,\eta(Z_{\Fe/\mathrm{H}})
\,p_{\mathrm{delay}}(t_0-t')
\,\dd Z_{\Fe/\mathrm{H}}\,\dd t',
\label{eq:event_rate_general}
\end{equation}
where $Z_{\Fe/\mathrm{H}}$ denotes the metallicity probed by iron abundance, $f_{\rm SFR}(Z_{\Fe/\mathrm{H}},t')$ is the iron-dependent cosmic \acrshort{sfh}, $\eta(Z_{\Fe/\mathrm{H}})$ is the number of merging systems formed per unit stellar mass, and $p_{\mathrm{delay}}$ is the delay-time distribution between progenitor star formation and \acrlong{bh} binary merger. The metallicity-dependent formation efficiency is described by~\cref{eq:metallicity_efficiency}.
For the delay-time distribution, we adopted a conventional power law
\begin{equation}
p_{\mathrm{delay}}(\tau)
=
\begin{cases}
A\,\tau^{-1}, & \tau_{\mathrm{min}} \leq \tau \leq t_{\mathrm{H}},\\
0, & \tau < \tau_{\mathrm{min}},
\end{cases}
\label{eq:dtd_power_law}
\end{equation}
where $A$ normalizes the distribution to unity between $\tau_{\mathrm{min}}$ and the Hubble time.
We considered $\tau_{\mathrm{min}}=10\,\mathrm{Myr}$ for the short-delay case and $\tau_{\mathrm{min}}=1\,\mathrm{Gyr}$ for the long-delay case.

\end{appendix}

\end{document}